\documentclass[journal=tches,eprint]{iacrtrans}%

\usepackage[T1]{fontenc}
\usepackage{graphicx}
\usepackage{amsmath}
\usepackage{mathtools}
\usepackage{amsfonts}
\usepackage{comment}
\usepackage{xspace}
\usepackage[breaklinks,colorlinks=true, citecolor=blue]{hyperref}
\usepackage[ruled,vlined,linesnumbered]{algorithm2e}
\SetAlFnt{\footnotesize}       
\SetAlCapFnt{\footnotesize}      
\SetAlCapNameFnt{\small}  
\usepackage{cleveref}
\usepackage{multirow}
\usepackage[colorinlistoftodos]{todonotes}

\usepackage{xcolor}
\usepackage{listings}
\usepackage{caption}
\usepackage{booktabs}
\usepackage[section]{placeins}

\usepackage{tikz-cd}
\usepackage{subcaption}
\usepackage{siunitx}

\makeatletter
\newcommand{\codlinelabel}[1]{%
  \edef\@currentlabel{\thelstnumber}
  \label{#1}%
}
\definecolor{codebg}{RGB}{248,248,250}
\definecolor{codenums}{RGB}{150,150,170}
\definecolor{codeframe}{RGB}{220,220,235}
\definecolor{codekw}{RGB}{0,73,138}
\definecolor{codestr}{RGB}{153,0,102}
\definecolor{codecm}{RGB}{90,110,120}

\definecolor{cmprimebg}{HTML}{D4E6F8}  
\definecolor{cmloopbg}{HTML}{FADDB8}  
\newcommand{\cmprime}[1]{{\setlength{\fboxsep}{1.5pt}\colorbox{cmprimebg}{#1}}}
\newcommand{\cmloop}[1]{{\setlength{\fboxsep}{1.5pt}\colorbox{cmloopbg}{#1}}}

\lstdefinestyle{c-elegant}{
  language=C,
  basicstyle=\ttfamily\small,
  keywordstyle=\color{codekw}\bfseries,
  stringstyle=\color{codestr},
  commentstyle=\itshape\color{codecm},
  numberstyle=\color{codenums}\footnotesize,
  backgroundcolor=\color{codebg},
  frame=single,
  rulecolor=\color{codeframe},
  frameround=tttt,
  breaklines=true,
  tabsize=2,
  showstringspaces=false,
  numbers=left,
  numbersep=8pt,
  xleftmargin=1.2em,
  framexleftmargin=1.2em,
  keepspaces=true
}

\crefname{lstlisting}{Listing}{Listings}
\Crefname{lstlisting}{Listing}{Listings}

\newcommand\Fp{\mathbb{F}_p}
\newcommand\Fpsq{\mathbb{F}_{p^2}}
\newcommand\Z{\mathbb{Z}}
\newcommand\Ep{\ensuremath{E^+}}
\newcommand\Em{\ensuremath{E^-}}

\newcommand{\fl}{\mathfrak{l}}

\newcommand{\csidh}{CSIDH\xspace}
\newcommand{\velu}{V\'elu's }

\SetKwFunction{evaliso}{Eval-$\ell_i$-isogeny}
\SetKwFunction{evalfiso}{Eval-$4$-isogeny}
\SetKwFunction{cswap}{Cswap}
\SetKwFunction{Isog}{IsoG}
\SetKwFunction{Mul}{Mul}
\SetKwFunction{IsoFix}{IsoFix}

\DeclareSIUnit{\bit}{bit}
\DeclareSIUnit{\bits}{bits}

\newcommand{\cl}{\mathrm{cl}}
\begin{document}
\title{From if to Oops: \\Turning a conditional into a key-recovery oracle}
%
%
\author{Gustavo Banegas \inst{1} \and Nia Dimitrova \inst{2} \and  Monika Trimoska \inst{2} }
\authorrunning{Banegas, Dimitrova, Trimoska}

\institute{
{Inria and Laboratoire d’Informatique de l’Ecole polytechnique, \\
Institut Polytechnique de Paris, Palaiseau, France}
\and
  {Eindhoven University of Technology, The Netherlands\\
    \email{gustavo@cryptme.in, nia.r.dimitrova@gmail.com, m.trimoska@tue.nl}}
}
\maketitle              
\begin{abstract}
Deterministic variants of CSIDH remove randomness from the group action, either through implementation techniques or, as in Houben's deterministic CSIDH and OSIDH-LD, by design. Since earlier fault attacks on CSIDH mostly target random point sampling, determinism seems to close that avenue. We show that it opens another: every secret-dependent step occurs at a fixed, public point of every execution, so an attacker can fault the same step repeatedly. Our attack forces the mask that selects which of two lanes computes the next isogeny to zero, for example by skipping one instruction. Whether the output changes reveals whether a secret exponent reaches a chosen threshold, and a binary search over these thresholds recovers the full key with one fault per execution: at most 297 oracle queries for OSIDH-LD and 222 for deterministic CSIDH with CSIDH-512-sized keys. We recover full OSIDH-LD keys in a software fault simulation of the reference implementation, and show on a ChipWhisperer-Lite that the targeted instruction is skipped with probability at least $0.94$ at an isolated fault position. Finally, we argue that fault resistance requires giving up determinism, and propose countermeasures relying on randomization.

\end{abstract}

\begingroup
\makeatletter
\def\@thefnmark{$\*$}\relax
\@footnotetext{\relax
Author list in alphabetical order; see
\url{https://ams.org/profession/leaders/CultureStatement04.pdf}.

This work was supported by:
\begin{itemize}
  \item the EuroTech Visiting Researcher Program 2025.
  \item the Dutch government through the 
  NWO XL project OCENW.XL.23.112 (ESCAPE).
  \item the HYPERFORM consortium, funded by France
  through Bpifrance, and by the France 2030 program under
  grant agreement ANR-22-PETQ-0008 PQ-TLS.
\end{itemize}
\def\ymdtoday{\leavevmode\hbox{\the\year-\twodigits\month-\twodigits\day}}
\def\twodigits#1{\ifnum#1<10 0\fi\the#1}

Date of this document: \ymdtoday.
}
\endgroup

\section{Introduction}
\label{sec:intro}
In recent years, there has been a boom in isogeny-based cryptography
development, from the algorithmic level to the implementation. One
scheme that has advanced in several aspects is CSIDH~\cite{Castryck0MPR18}.
Since its proposal in 2018, there have been several developments,
such as constant-time implementations~\cite{MeyerCR19,OnukiAYT20,AdjCR20,Cervantes-Vazquez19,Chavez-SaabCJR20,Chi-DominguezR22}.
A further milestone was CTIDH~\cite{BanegasBCCLMSS21}, which
improved upon the previous constant-time implementations. More
recently, there have been attempts to develop CSIDH implementations in a deterministic way,
such as: dCSIDH~\cite{secsidh}, dCTIDH~\cite{CamposHMR25}, and a
hardened version of CTIDH~\cite{BanegasHS25}, which is deterministic
and free of dummy operations.

However, all the previous works were engineering efforts; that is,
the changes were not at the algorithmic level, but rather
implementation techniques that allow the operations to be performed
in a deterministic way. To address this at the algorithmic level,
Houben proposed changes to CSIDH to make it fully
deterministic~\cite{C:Houben25}. Moreover, he also presented another version that
uses the idea of orientations on curves, making this version
deterministic as well~\cite{Houben25_asiacrypt}. Subsequently, Wang, Lai, 
Lin, and Zhao~\cite{OSIDH26} implemented the scheme 
in~\cite{Houben25_asiacrypt} using parallel techniques to speed up.

Implementations on embedded devices must also be secure against fault attacks,
and CSIDH has been studied extensively in this respect. Existing attacks
target the two sources of non-determinism above: dummy isogenies~\cite{CamposKMOS20,CamposKM21,legrow_20,legrow_24},
and random point sampling~\cite{EC:BKLMPR23}. \emph{Neither is
present in Houben's deterministic CSIDH or in OSIDH-LD: there are no
dummy operations, and the kernel points are part of the public
parameters.}

This seems to close the fault avenue, but the secret still decides, at
every step, which of the two trails advances. In constant-time code, this
decision is a conditional swap driven by a secret mask, which is invisible
to timing but not to faults. Because the schedule is fixed and public, the
attacker can fault the same decision in every execution: determinism makes
the fault reproducible.

\paragraph{Related work.} 
\label{sec:related}
LeGrow and Hutchinson~\cite{legrow_20} provide a
\emph{mathematical} analysis of fault attacks against constant-time CSIDH
implementations that use dummy isogenies. They show that if the real isogenies
of each prime degree are executed before the dummy ones
(\emph{real-then-dummy}), an attacker can recover the entire private key up to
sign with absolute certainty using a \emph{binary search} strategy that
requires only
\(\sum_j (\lceil \log_2(b_j) \rceil + 1)\) carefully timed faults, for $\mathbf{b}$ a bound vector defining the keyspace.

Chiu, LeGrow, and Xiong~\cite{legrow_24} bridge the gap between the 
theory and \emph{practice}. They build a complete fault-injection attack
pipeline against \emph{real-then-dummy} CSIDH running on STM32
microcontrollers, using low-cost voltage glitches. Their work highlights
practical challenges ignored by earlier models---such as non-deterministic
fault success rates, device resets, and the need to tune glitch timing and
duration per isogeny degree. They experimentally validate the binary-search
attack, propose new bounds on the number of faults required under realistic
success probabilities, and are the first to empirically test the
\emph{dynamic random decision-vector} countermeasure.

In~\cite{EC:BKLMPR23}, the authors introduce the notion of a
\emph{disorientation fault}: a perturbation that causes one step of the
group-action computation to proceed in the opposite direction from what
the secret key prescribes.
When an $\ell_i$-isogeny step is misdirected in this way, the output
curve lies $\ell_i^2$-isogenies away from the intended image, and
comparing the faulty output with the correct one yields a one-bit oracle
on the affected exponent.
In standard CSIDH implementations, the probabilistic point-sampling step
is the natural fault site: corrupting the square-root test can force the
sampled point onto the quadratic twist, reversing the sign of the
computed isogeny.

\textbf{Our attack} is likewise a disorientation fault, but it targets a
\emph{dummy-free} deterministic implementation where no point sampling
occurs at all: the kernel points $P$ and $Q$ are fixed as part of the
public parameters, so there is no sampling step to corrupt.
Instead, the fault corrupts the secret-dependent lane-selection mask,
causing one isogeny to be computed on the wrong trail.
In this sense, our work extends the disorientation paradigm to a setting
that gives a faulty oracle for free, and it
reinforces the broader lesson of~\cite{EC:BKLMPR23}: determinism is the
key property that makes a fault target reproducible.

\paragraph{Contributions.}
We show that removing randomness from the CSIDH group action, as
Houben's deterministic CSIDH~\cite{C:Houben25} and
OSIDH-LD~\cite{OSIDH26} do by design, \emph{enables a simple fault attack}.
\textbf{The weakness lies in the fixed, public schedule, not in a particular
implementation.} More precisely:
\begin{itemize}
  \item \textbf{Safe-error oracle.} Both schemes select the lane of each
  isogeny step with a constant-time conditional swap whose mask depends on
  the secret. Forcing this mask to zero, e.g., by skipping one instruction,
  changes the output if and only if the step index $j$ exceeds $s_i$. Since
  the schedule is public, the same step can be faulted in every execution.

  \item \textbf{Key recovery.} A binary search per secret component recovers
  the key in at most $n\lceil\log_2(B+1)\rceil$ queries: 222 for
  deterministic \csidh ($n = 74$, $B = 5$). For OSIDH-LD (\texttt{ld256}), a
  two-stage search handles the cycle structure and the torsion powers
  $e_i > 1$ in at most 297 queries, matching the information-theoretic bound.

  \item \textbf{Experiments.} We recover full keys in a software fault
  simulation of the OSIDH-LD reference implementation~\cite{OSIDH26}. On a
  ChipWhisperer-Lite, a clock glitch aligned to the isolated mask
  computation skips the instruction in 50 of 50 attempts ($p \geq 0.94$ at
  95\% confidence).

  \item \textbf{Countermeasures.} We randomize the order of the primes and of
  the outer loop, keeping both schemes dummy-free. For OSIDH-LD, shuffling
  plus one extra all-zero iteration turns every effective fault into an
  abort that hides the targeted prime, provided the lane check is enforced.
  For deterministic \csidh, where a fault yields a valid but wrong curve,
  shuffling raises the cost to 723--1605 successful faults, and randomizing
  the outer loop leaves only a statistical estimate of the key.
\end{itemize}

Table~\ref{tab:overview} summarizes the attack and the countermeasures.

\begin{table}[!htb]
  \centering
  \small
  \caption{Fault complexity and leakage of our attack on each variant.
  \csidh uses the CSIDH-512 primes ($n = 74$, $B = 5$); OSIDH-LD uses
  \texttt{ld256} ($n = 49$, $B = 35$, $r = 13$), with $s_i \in [0, Be_i]$.
  \#\,Faults: worst-case oracle queries for the deterministic variants;
  expected successful faults for the randomized ones (best and worst case
  for \csidh). For OSIDH-LD with prime-order randomization, the attacker
  learns the indices $i$ whose $s_i$ lies in the top range
  $((B-1)e_i, Be_i]$, which fixes $s_i = B$ when $e_i = 1$. Added cost:
  extra isogeny computations over the unprotected scheme.
  n.d.: not determined.}
  \label{tab:overview}
  \begin{tabular}{@{}llrlc@{}}
    \toprule
    Variant & Scheme & \#\,Faults & Leakage & Added cost \\
    \midrule
    Deterministic & \csidh   & 222 & full key & -- \\
                  & OSIDH-LD & 297 & full key & -- \\
    \addlinespace
    Prime-order randomization & \csidh   & 723--1605 & key candidate             & 0 \\
                              & OSIDH-LD & 219       & $\{i : s_i > (B-1)e_i\}$  & 0 \\
    \addlinespace
    \quad + outer-loop randomization & \csidh   & n.d. & statistical estimate & 0 \\
    \quad + one extra iteration      & OSIDH-LD & --   & --                   & $+1/B$ ($\approx 2.9\%$) \\
    \bottomrule
  \end{tabular}
\end{table}

Our code is available at: \href{https://github.com/gbanegas/faulty\_isogenies}{https://github.com/gbanegas/faulty\_isogenies}.


\section{Background}
\label{sec:background}
This section gives a brief mathematical background necessary to 
understand the rest of the paper.
\paragraph{Elliptic curves, twists, and isogenies.}
In the \csidh setting, we work with supersingular Montgomery curves over
$\Fp$, written as
\[
    E_A:\ y^2 = x^3 + A x^2 + x,
\]
with $A \in \Fp$ and $A^2 \neq 4$.
Their quadratic twist is
\(    E_{-A}:\ y^2 = x^3 - A x^2 + x.\)
The curves $E_A$ and $E_{-A}$ are isomorphic over $\mathbb{F}_{p^2}$ but,
in general, not over $\Fp$.
In the \csidh parameter regime, where $p \equiv 3 \bmod 4$, supersingular
curves defined over $\Fp$ satisfy $\#E_A(\Fp)=p+1$.
The curve $E_0$ with $j(E_0)=1728$ is exceptional in that it is
$\Fp$-isomorphic to its quadratic twist, and it is used as the base point
of the group action.

An isogeny $\phi:E\to E'$ is a non-constant morphism of elliptic curves
that preserves the group law.
It is uniquely determined by its kernel, a finite subgroup of $E$, and its
degree equals the size of that kernel.
In \csidh, the relevant isogenies have small odd prime degree $\ell_i$ and
connect vertices of the supersingular isogeny graph over $\Fp$.

\paragraph{\csidh and the class-group action.}
Let
\[
    p = (4 \cdot \prod_{i=1}^n \ell_i) - 1,
\]
where the $\ell_i$ are distinct small odd primes.
This choice gives access to efficient $\ell_i$-isogeny computations. This geometric setup has a surprising correspondence in the algebraic world. 
For the imaginary quadratic order
\[
    \mathcal{O} = \mathbb{Z}[\pi] \cong \mathbb{Z}[\sqrt{-p}],
\]
the ideal class group $\cl(\mathcal{O})$ acts freely and transitively on
the set $\mathcal{E}$ of supersingular elliptic curves over $\Fp$ with
$\Fp$-endomorphism ring isomorphic to $\mathcal{O}$.
We denote the action of an ideal class $[\mathfrak{a}]$ on
$E \in \mathcal{E}$ by $\mathfrak{a} * E$.

Each prime $\ell_i$ splits in $\mathcal{O}$ as
\[
    (\ell_i) = \mathfrak{l}_i \bar{\mathfrak{l}}_i,
\]
and the corresponding ideal classes
$[\mathfrak{l}_i]$ and $[\mathfrak{l}_i]^{-1}$ represent the two possible
directions of an $\ell_i$-step in the graph.
Accordingly, a private key is an exponent vector
\[
    (e_1,\dots,e_n)\in[-B,B]^n,
\]
representing the class
\[
    [\mathfrak{a}] = \prod_{i=1}^n [\mathfrak{l}_i]^{e_i}.
\]
Applying $[\mathfrak{a}]$ to the base curve $E_0$ amounts to walking in the
isogeny graph, taking a positive $\ell_i$-isogeny when $e_i>0$ and a
negative $\ell_i$-isogeny when $e_i<0$.
The resulting curve $E'=\mathfrak{a}*E_0$ is the public key.
Since the action is commutative,
\[
    \mathfrak{a} * (\mathfrak{b} * E_0)
    = \mathfrak{b} * (\mathfrak{a} * E_0),
\]
two parties obtain the same shared secret by applying their secret classes
in either order.

We use the following shorthand throughout the paper.
For a set $S$ of primes $\ell_i$, we write $\varphi_S$ for the isogeny
whose kernel is generated by the corresponding $\ell_i$-torsion points, and
$\widehat{\varphi_S}$ for its dual.
For an ideal class $[\mathfrak{a}]$, we write $\varphi_{\mathfrak{a}}$ for
an isogeny realising the action of $[\mathfrak{a}]$.

\paragraph{Point sampling and $x$-only arithmetic.}
A key feature of \csidh is that the action can be implemented efficiently
on Montgomery curves using only $x$-coordinates.
This enables the use of $x$-only arithmetic together with \velu formulas for
the computation and evaluation of isogenies.
Given a point generating a kernel of order $\ell$, Vélu's formulas produce
the corresponding $\ell$-isogeny, and the cost is essentially linear in the
kernel size.

The sign of a step determines on which curve the kernel point must lie.
To compute a positive $\ell_i$-isogeny from $E_A$, one needs an
$\Fp$-rational point of order $\ell_i$ on $E_A$.
To compute a negative $\ell_i$-isogeny, one instead needs a point of order
$\ell_i$ on the quadratic twist $E_{-A}$.
Standard \csidh obtains such points by \emph{point sampling}: one chooses an
$x$-coordinate at random and tests whether $x^3 + A x^2 + x$ is a square or
a non-square in $\Fp$.
This determines whether the sampled point lies on $E_A$ or on its twist.
After projecting to the appropriate $\ell_i$-torsion component, one obtains
a kernel generator for the desired isogeny.

To reduce the number of samplings, implementations reuse sampled points by
\emph{pushing} them through successive isogenies.
If $\varphi$ is an isogeny with kernel generated by $R$, then $\varphi(P)$
is a point on the image curve whose order is reduced by the torsion already
consumed in the step.
As a consequence, fresh point sampling must be performed repeatedly during a
full group-action computation.

\paragraph{Secure implementations.}
For this paper, a constant-time implementation is one whose control flow,
memory accesses, and instruction sequence do not reveal secret-dependent
information through timing.
This does not mean that every execution has exactly the same runtime in an
absolute sense; rather, it means that the variations are independent of the
secret key.
This distinction is important in the \csidh setting: even when timing
leakage is removed, the implementation must still realise the secret-dependent
choice between positive and negative isogeny steps.
Our attack surface comes from how that choice is encoded and manipulated in
deterministic implementations.

\section{Deterministic CSIDH variants}
\label{sec:dvariants}
\subsection{Deterministic CSIDH}
\label{sec:dcsidh}

In~\cite{C:Houben25}, the idea is explained in full generality and it
is used to develop two algorithms that differ in the number of
\emph{kernel points} used. We focus here only on the concrete
application to the \csidh setting, as this is the one that we
subsequently show an attack for. The attack ideas transfer, at least
at a structural level, to the broader deterministic framework
of~\cite{Houben25_asiacrypt}: both schemes share the same fundamental
mechanism of a secret-dependent lane-selection mask implemented as a
constant-time conditional swap, and it is this mechanism---not any
detail specific to the \csidh group action---that our fault exploits.
We return to this point in \Cref{sec:attack:osidh}, where we show
that the oracle-based key recovery applies to OSIDH-LD with only
minor adaptations to account for the cycle structure and the torsion
powers $e_i > 1$.

The norm of the endomorphism $\pi - 1$ as an algebraic integer is
$p+1$. This implies that the principal $\mathcal{O}$-ideal generated
by $\pi - 1$ factors into
\[
    \mathfrak{c} \cdot \prod_{i=1}^n \fl_i,
\]
where $\mathfrak{c}$ denotes the ideal $(4, \pi -1)$ corresponding to
the 4-isogeny. This yields the following class group relation, proven
independently in both~\cite[Lemma~8]{EC:CasPanVer20}
and~\cite{IWSEC:OnuTak20}:
\[
    [\mathfrak{c}]^{-1} = [\fl_1] \cdots [\fl_n].
\]
Hence, to compute the group action corresponding to the exponent
vector $(1,\dots,1)$, one can alternatively compute a negative
4-isogeny. In other words, $\pi-1$ as an endomorphism can be
decomposed as $\varphi_{S_1} \circ \varphi_{S_2} \circ
\varphi_{\{4\}}$, for $S_1$ and $S_2$ two disjoint sets together
containing all primes $\ell_i$.

Restricting to exponent vectors in $[0,1]$, for a given vector
$(e_1,\dots,e_n)$ corresponding to the action $\mathfrak{a} \ast E$,
we define
\[
S_1 = \{\ell_i \mid e_i = 1\} \quad \text{and} \quad
S_2 = \{\ell_i \mid e_i = 0\},
\]
so that both $\varphi_{S_1}$ and $\widehat{\varphi_{S_2}} \circ
\widehat{\varphi_{\{4\}}}$ map $E$ to the same curve $E' = \mathfrak{a}
\ast E$. Writing $\varphi^+$ and $\varphi^-$ for $\varphi_{S_1}$ and
$\widehat{\varphi_{S_2}}$ respectively, the core idea of the
deterministic variant is to compute \emph{both} trails simultaneously.
In this way, the number and degree of isogenies computed is the same
regardless of the private key instance, and since computing a positive
or negative $\ell_i$-isogeny has the same runtime, a passive
attacker\footnote{An attacker who can perform side-channel analysis but
cannot alter execution via fault injection.} cannot detect which trail
was advanced at each step. Point sampling is avoided by appending
kernel points $P \in E(\Fp)$ and $Q \in E^t(\Fp)$, both of order
$\prod_i \ell_i$, to the public key, so that both trails can be
computed by the standard point-pushing strategy. This is detailed in
\Cref{alg:dcsidh_marc} (same as~\cite[Algorithm~3]{C:Houben25}), and
generalised to compute the entire group action by repeating the
procedure $B$ times.

\begin{algorithm}[htp]
\caption{Deterministic CSIDH}
\label{alg:dcsidh_marc}
\KwIn{An elliptic curve $E/\Fp$, points $P \in E(\Fp)$ and
$Q \in E^t(\Fp)$ of maximal odd order, an integer $B \in \Z_{>0}$,
a vector of integers $(s_1,\ldots,s_n) \in [0,B]^n$.}
\KwOut{The curve $E' := \prod_{i} \mathfrak{l}_i^{\,s_i} \ast E$,
points $P' \in E'(\Fp)$, $Q' \in E'^{t}(\Fp)$ of maximal odd order.}

$(P', Q') \gets (P, Q)$\;
\For{$j \gets 1$ \KwTo $B$}{
    $(P^+, P^-) \gets (P', Q')$\;
    $m \gets \prod_i \ell_i$\;
    \For{$i \gets 1$ \KwTo $n$}{
        \eIf{$j \leq s_i$}{
            $m \gets m/\ell_i$\;
            $K \gets [m]P^+$\;
            $(\Ep, P^+, Q') \gets \evaliso(\Ep, K, P^+, Q')$\;
            $P^- \gets [\ell_i]P^-$\;
        }{
            $m \gets m/\ell_i$\;
            $K \gets [m]P^-$\;
            $(\Em, P^-, P') \gets \evaliso(\Em, K, P^-, P')$\;
            $P^+ \gets [\ell_i]P^+$\;
        }
    }
    $(\Em, P') \gets \evalfiso(\Em, P')$\;
    \textbf{assert} $\Ep = \Em$\;
}
\Return{$(E^+, P', Q')$}\;
\end{algorithm}

To avoid resampling kernel points after each iteration, $P$ and $Q$
are pushed through the opposite trails. Since $\langle P \rangle \cap
\langle Q \rangle = \{0\}$, no torsion is lost in the process:
$\varphi^-(P)$ is a maximal odd order $\Fp$-rational point on $E'$,
and $\varphi^+(Q)$ is a maximal odd order point that is $\Fp$-rational
on $E'^t$. The $\evalfiso$ step computes a negative 4-isogeny using
explicit formulas to bring the two trails together, and the
if-branch is implemented as a $\cswap$---a constant-time conditional
swap of two memory registers.

\subsection{OSIDH-LD}
\label{sec:osidh}

The scheme of~\cite{Houben25_asiacrypt} generalises the deterministic
group-action framework of \Cref{alg:dcsidh_marc} to supersingular
curves equipped with orientations by imaginary quadratic orders of
large discriminant. The resulting protocol, called \emph{OSIDH-LD} in
the accompanying implementation paper~\cite{OSIDH26}, retains the four
properties that motivated our study: it is fully deterministic,
dummy-free, constant-time, and branch-free.

The state carried through the computation is a \emph{full kernel
representation} of the orientation. The orientation is chosen to be a separable endomorphism with a cyclic kernel
and degree $M^r$ for $M=\prod_{i=1}^n \ell_i^{e_i}$, hence it can be represented as a composition of $r$ isogenies $\varphi_j$, each of degree $M$ and a kernel generated by $P_j \in E(\Fp)$. This is a key component that allows for the use of only $\Fpsq$-arithmetic by keeping track of the kernels of all isogenies in the composition and their duals. A full kernel representation is thus a cycle of $r$ tuples
$(E_j, P_j, Q_j)$, where each $E_j$ is a supersingular curve over
$\mathbb{F}_{p^2}$ and $P_j, Q_j$ are torsion points of maximal odd
order $M$. 
The group action is evaluated by running $B$
iterations of a core loop that processes all $r$ tuples and all prime
factors of $M$ in a fixed, publicly known order. For the
\texttt{ld256} parameter set, the concrete values are
\[
  r = \mathtt{CYCLE\_LENGTH} = 13,\qquad
  B = \mathtt{ITERATIONS} = 35,\qquad
\]
\[
 n = \mathtt{P\_LEN} + \mathtt{M\_LEN} = 49.
\]

The torsion powers are $e_0 = 4$ for prime $3$, $e_3 = 2$ for prime
$11$, and $e_i = 1$ for the remaining $47$ primes, so the secret key
components satisfy $s_i \in [0, B \cdot e_i]$.

Algorithm~\ref{alg:osidh_ld} describes one iteration of the group-action
evaluation (same as~\cite[Algorithm~1]{Houben25_asiacrypt}).  The full
group action is obtained by running $B$ successive applications of
this algorithm, each time feeding the output orientation cycle as the
next input.  Comparing with Algorithm~\ref{alg:dcsidh_marc}, the key
structural differences are fourfold.  First, the state is a
\emph{cycle} of $r$ tuples rather than a single curve with two
kernel points; the outer loop over $j$ processes each tuple
independently, which is precisely the parallelism exploited
in~\cite{OSIDH26}.  Second, because the torsion order $M$ may have
prime-power factors (i.e.\ $e_i > 1$), each prime $\ell_i$ requires
$e_i$ inner steps, each guarded by the same $\cswap$ pattern.  Third,
there is no closing 4-isogeny step; instead, an isomorphism
identification is applied at the end of the $j$-loop to ensure that
the two lanes $R_0$ and $R_1$ share the same curve
coefficients~\cite[Remark~1]{OSIDH26}. Finally, the positive $\varphi_j^+$ and negative $\varphi_j^-$ sides are not computed starting from the same curve but from neighbouring ones, which has an (insignificant) impact on the form of the disorientation fault we obtain. Namely, disorienting an $\ell_i$-isogeny results in a faulty curve that is an $\ell_i$-isogeny away from the correct one, instead of $\ell_i^2$ as in the case of deterministic \csidh.

\begin{algorithm}[htp]
\caption{OSIDH-LD group action (one iteration)}
\label{alg:osidh_ld}
\KwIn{A full kernel representation
$\mathcal{E} = \{(E_j, P_j, Q_j) \mid 1 \leq j \leq r\}$ of a
$\mathbb{Z}[\sigma]$-orientation on $E$, a vector
$(s_1,\ldots,s_n) \in \prod_{i=1}^{n}[0, e_i]$.}
\KwOut{A full kernel representation
$\mathcal{E}' = \{(E'_j, P'_j, Q'_j) \mid 1 \leq j \leq r\}$ of the
$\mathbb{Z}[\sigma]$-orientation on
$E' = \bigl[\prod_{i=1}^{n} \mathfrak{l}_i^{\,s_i}\bigr] \ast E$.}

\For{$j \gets 1$ \KwTo $r$}{
    $R_0 \gets (E_j,\; P_j,\; Q_j)$\;
    $R_1 \gets (E_{j+1},\; Q_{j+1},\; P_{j+1})$\tcp*{indices mod $r$}
    $m   \gets \prod_{i=1}^{n} \ell_i^{\,e_i}$\;
    \For{$i \gets 1$ \KwTo $n$}{
        \For{$t \gets 1$ \KwTo $e_i$}{
            $b \gets (t > s_i)$\tcp*{secret-dependent control bit}
            $\cswap(R_0,\; R_1,\; b)$\;
            $m \gets m / \ell_i$\;
            $K \gets \Mul(R_0[0],\; R_0[1],\; m)$\;
            $(R_0[0],\; \{R_0[1], R_0[2]\})
              \gets \Isog(R_0[0],\; \{R_0[1], R_0[2]\},\; K,\; \ell_i)$\;
            $R_1[1] \gets \Mul(R_1[0],\; R_1[1],\; \ell_i)$\;
            $\cswap(R_0,\; R_1,\; b)$\;
        }
    }
    $R_0[1] \gets R_1[2]$\;
    $E'_j \gets \IsoFix(R_0,\; R_1)$\tcp*{isomorphism identification for $E'_j$}
}
\Return{$E'_1, \ldots, E'_r$}\;
\end{algorithm}

The $\Mul$ and $\Isog$ calls are the same scalar-multiplication and
isogeny-evaluation primitives as in Algorithm~\ref{alg:dcsidh_marc}; 
$\IsoFix$ denotes the isomorphism step that reconciles the two lanes after each
$j$-iteration. The $\cswap$ on line~8 of the inner-most loop is
exactly the operation whose secret-dependent mask is our fault target,
as detailed below.

The feature that matters for our attack is not the presence or
absence of explicit branches, but whether the computation performs
\emph{secret-dependent routing} of the active lane in a deterministic
schedule. OSIDH-LD does exactly that, via masked swaps. In the
reference implementation, the function \texttt{eval\_action} follows the same 
structural pattern as Algorithm~\ref{alg:dcsidh_marc}: it computes a 
bitmask from the secret thresholds, uses that mask to swap the 
two lanes, performs the isogeny work on the active lane, and 
then swaps the lanes back, the code that implement this part
is in~Listing~\ref{lst:osidh-fault-site}, where \texttt{compare\_gt} 
returns the all-ones word when
$t \leq s_i$ and zero otherwise, and \texttt{swap\_vertices} performs
a branchless conditional swap on all five field elements of each
\texttt{vertex\_t}. When $t \leq s_i$, the mask is all-ones and the
active lane is moved into position \texttt{R[0]}; when $t > s_i$, the
mask is zero and no swap occurs.
\begin{lstlisting}[  basicstyle=\ttfamily\scriptsize, numberstyle=\tiny, numbersep=4pt,
  aboveskip=4pt, belowskip=0pt,
  caption={Fault target in \texttt{eval\_action} (\texttt{protocols.c}).},
  label={lst:osidh-fault-site}]
digit_t b = compare_gt(t, s[i]); 
swap_vertices(&(R[0]), &(R[1]), b);
/* ... scalar multiplication + isogeny evaluation on R[0] ... */
swap_vertices(&(R[0]), &(R[1]), b);
\end{lstlisting}

This is the exact analogue of the \texttt{bool~c} assignment in
Listing~\ref{lst:isog-step}. A fault that forces the mask to $0$ when it
should be all-ones leaves the wrong lane active and executes one
isogeny step in the wrong direction---the same mechanism exploited in
Section~\ref{sec:attack}. Crucially, this observation shows that
\emph{branch-free code is not automatically fault-resistant}: the
attack surface is the secret-dependent selection mask, regardless of
whether the implementation contains an explicit \texttt{if} statement.

\section{Attack}
\label{sec:attack}
In this section, we present the core idea of the fault attack and
identify the computation that an attacker must fault in the deterministic
CSIDH variant of~\cite{C:Houben25}.
For the attack, the relevant objects are the two companion states
$\Ep$ and $\Em$ together with the associated points
$P^+$ and $P^-$. At each iteration, the implementation uses a
secret-dependent control bit to decide which lane is active, performs
the isogeny computation on that lane, and then swaps the two lanes
back. Our attack targets exactly this control value.

\subsection{Attack Model}
\label{sec:attack:model}
We consider an active attacker who can query the target device multiple
times while the same long-term secret key is used, and who can inject a
single transient fault during one chosen execution. This is the standard
setting for implementation-level fault attacks on isogeny-based
protocols, and it is consistent with prior work such
as~\cite{EC:BKLMPR23}.

The concrete fault model is deliberately weak. We assume that the
attacker can force one secret-dependent Boolean control value from
\texttt{True} to \texttt{False}, or equivalently from $1$ to $0$, at a
chosen step of the deterministic loop. Operationally, this models an
instruction skip, a corrupted comparison result, or a transient fault on
a register carrying the branch condition. We do \emph{not} assume the
ability to set arbitrary values, to modify the secret key directly, or
to inject multiple coordinated faults in one execution.

The second assumption is that the attacker can target a specific loop
iteration, and therefore a specific prime-degree isogeny step. This is
precisely where determinism matters: because the implementation executes
the prime factors in a public and reproducible order, the same step is
reached at the same relative time in every execution. As a result, once
the attacker has aligned the fault to one chosen event, the same event
can be targeted repeatedly.

Under these assumptions, a successful $1 \rightarrow 0$ fault causes the
implementation to execute one isogeny step on the wrong lane. The
fault-free and faulted executions then produce different outputs, which
turns the resulting comparison into a one-bit oracle on the
secret-dependent control value; \Cref{sec:results} shows that the fault
can be injected at the targeted control site.

A successful fault can have two outcomes depending on implementation choices and the fault injection place: producing a disorientation or an execution error. We stress that in our attack, both outcomes result in recovering the same amount of information, namely one exponent (or a more restricted bound in the case where $e_i>1$ or $B>1$) of the secret key.

\subsection{Key-recovery procedure}
\label{sec:attack:recovery}

We now show that the one-bit oracle suffices to recover the entire
secret key $(s_1,\ldots,s_n) \in [0,B]^n$.

\paragraph{The lane-selection mask.}
\label{par:mask}
We fix the convention once, as both schemes use it. The isogeny of a
step is always computed on the \emph{default} lane---$\Ep$ together with
$P^+$ in \Cref{alg:dcsidh_marc}, the register \texttt{Aright} in
Listing~\ref{lst:isog-step}---and the secret selects a lane by swapping
the two before the computation and swapping them back afterwards. Since
$s_i$ counts the steps that belong to the default lane, the mask is
\emph{set} exactly when the step counter has passed the threshold:
\begin{equation}\label{eq:mask}
  c = 1 \iff j > s_i .
\end{equation}
This is what Listing~\ref{lst:isog-step} computes as
\texttt{bool~c~=~(e[i]~<=~k)}, with $k$ counting from $0$ and $j = k+1$,
so that $s_i \leq k \iff j > s_i$. The branch condition of
\Cref{alg:dcsidh_marc} is the complement of~$c$; it is the mask, not the
branch condition, that the attacker faults.

\paragraph{Oracle definition.}
Fix a prime index $i \in \{1,\ldots,n\}$ and an outer-loop counter
$j \in \{1,\ldots,B\}$.  Let $\mathrm{PK}(\mathbf{s})$ denote the
output of \Cref{alg:dcsidh_marc} on secret key $\mathbf{s}$, and let
$\mathrm{PK}_{\mathsf{f}}^{(j,i)}(\mathbf{s})$ denote the output when
the mask at step $(j,i)$ is forced to~$0$. Define
\[
  \mathcal{O}(j,i)
  \;=\;
  \bigl[\mathrm{PK}(\mathbf{s}) =
        \mathrm{PK}_{\mathsf{f}}^{(j,i)}(\mathbf{s})\bigr].
\]
If $j \leq s_i$, the mask is already $0$ by \Cref{eq:mask}; forcing it
to $0$ has no effect and both outputs agree, so $\mathcal{O}(j,i)=1$.
If $j > s_i$, the mask is $1$; suppressing the swap leaves the step on
the default lane instead of the one the key prescribes, and the outputs
differ, so $\mathcal{O}(j,i)=0$. Hence
\begin{equation}\label{eq:oracle}
  \mathcal{O}(j,i) = 1 \iff j \leq s_i.
\end{equation}
The fault is thus a \emph{safe error}: the bit is read from whether
the fault changes the output, which turns a single skipped swap into a
reliable one-bit oracle.

\paragraph{Binary search.}
For a fixed~$i$, the map $j \mapsto \mathcal{O}(j,i)$ is monotone
non-increasing: it returns~$1$ for all $j \leq s_i$ and~$0$ for all
$j > s_i$.  The threshold $s_i$ is therefore the largest $j$ with
$\mathcal{O}(j,i) = 1$, and $s_i = 0$ exactly when
$\mathcal{O}(1,i) = 0$.  A standard binary search over
$j \in \{1,\ldots,B\}$ locates it using at most
$\lceil\log_2(B+1)\rceil$ oracle queries.  Repeating for each
coordinate~$i$ independently gives a total of
$n \cdot \lceil\log_2(B+1)\rceil$ oracle queries to recover the full key.

\paragraph{Concrete complexity.}
For the CSIDH-512 parameter set~\cite{Castryck0MPR18}, where $n = 74$
and $B = 5$, the binary search requires
$\lceil\log_2 6\rceil = 3$ queries per coordinate, yielding
$74 \times 3 = 222$ oracle queries in total ($197$ on average).
Each oracle query corresponds to one fault injection followed by an
output comparison. With the repeated-trial model and the rate
$p = 3/1281$ of \Cref{sec:results:dcsidh-chip}, a full recovery takes
about $6.9\times10^{5}$ glitch attempts, between $2.4\times10^{5}$ and
$3.4\times10^{6}$ across the Clopper--Pearson interval.

\subsection{Attacking OSIDH-LD}
\label{sec:attack:osidh}

The fault attack described above extends naturally to OSIDH-LD
(\Cref{sec:osidh}).  The structural argument is the same: the
implementation is fully deterministic, so every secret-dependent
branching decision occurs at a predictable point in each execution,
and a single $1 \rightarrow 0$ fault at that point misroutes one
isogeny step onto the wrong lane.  What differs from the dCSIDH
setting is the shape of the secret key space, the presence of a
cycle dimension, and the fact that two primes carry torsion power
$e_i > 1$.  We spell out these differences below.

\paragraph{Fault target.}
The sensitive assignment in \texttt{eval\_action}
(see \Cref{lst:osidh-fault-site}) is
$b \leftarrow \texttt{compare\_gt}(t, s[i])$, which appears inside a
double loop: an outer loop over prime index
$i \in \{1,\ldots,n\}$ and an inner loop over step $t \in
\{1,\ldots,e_i\}$, where $e_i = \mathtt{TORSION\_ODD\_POWERS[i]}$.
This structure is repeated for each of the $r =
\mathtt{CYCLE\_LENGTH} = 13$ orientation tuples in the cycle.
For the \texttt{ld256} parameter set, the prime slots are split
into $\mathtt{P\_LEN}$ primes with kernel in $E[p+1]$ and
$\mathtt{M\_LEN}$ primes with kernel in $E[p-1]$, giving
$n = \mathtt{P\_LEN} + \mathtt{M\_LEN} = 49$ independently
recoverable key components.

\paragraph{Oracle.}
At outer iteration $u$, the effective key value for component $i$ is
$s_{\mathit{eff}} = s_i - u \cdot e_i$, clamped to $[0,e_i]$, and
\Cref{eq:mask} reads $b = 1 \iff t > s_{\mathit{eff}}$: the first
$s_{\mathit{eff}}$ of the $e_i$ inner steps run on $R_0$ and the rest
are routed to $R_1$.
Let $\mathsf{PK}(\mathbf{s})$ denote the unfaulted public key and
$\mathsf{PK}^{(u,i,t)}_{\mathsf{f}}(\mathbf{s})$ the output when
\texttt{compare\_gt} is forced to $0$ at outer iteration $u$, prime
index $i$, and inner step $t$. The oracle is
\[
  \mathcal{O}(u,i,t)
  = \bigl[\mathsf{PK}(\mathbf{s})
    = \mathsf{PK}^{(u,i,t)}_{\mathsf{f}}(\mathbf{s})\bigr].
\]
The public-key comparison checks \emph{all} $r = 13$ $j$-invariants of
the output orientation cycle, so any misdirected isogeny step
produces a detectable divergence.  At $t = 1$ the mask is set precisely
when $s_{\mathit{eff}} = 0$, so by the same argument as in
\Cref{eq:oracle},
\[
  \mathcal{O}(u,i,1) = 1 \iff s_i > u \cdot e_i.
\]

\paragraph{Two-stage key recovery.}
For a fixed prime index~$i$, a binary search over $u \in \{0,\ldots,B\}$
with $t = 1$ locates the smallest $u^{*}$ with
$\mathcal{O}(u^{*},i,1) = 0$, using at most
$\lceil\log_2(B+1)\rceil = 6$ queries. Together with
$\mathcal{O}(u^{*}-1,i,1) = 1$ this gives
$s_i \in ((u^{*}-1)\cdot e_i,\; u^{*}\cdot e_i]$.

For primes with $e_i = 1$ (the remaining $47$ primes in \texttt{ld256})
the block is a single value and the search ends there, with
$s_i = u^{*}$. For the two primes with $e_i > 1$ (prime $3$ with
$e_1 = 4$ and prime $11$ with $e_4 = 2$) a second stage is needed: a
binary search over $t \in \{1,\ldots,e_i\}$ at outer iteration
$u^{*}-1$, where the mask at step $t$ is set exactly when
$t > s_i - (u^{*}-1)\cdot e_i$. The threshold is the largest $t$ with
$\mathcal{O}(u^{*}-1,i,t) = 1$, which resolves $s_i$ within the block
using at most $\lceil\log_2 e_i\rceil$ additional queries.

\paragraph{Concrete complexity.}
For the \texttt{ld256} parameter set
($B = \mathtt{ITERATIONS} = 35$, $n = 49$), the total oracle query
count is
\[
  47 \cdot 6
  \;+\; \bigl(6 + \lceil\log_2 4\rceil\bigr)
  \;+\; \bigl(6 + \lceil\log_2 2\rceil\bigr)
  \;=\; 282 + 8 + 7 \;=\; 297 \text{ oracle queries,}
\]
and $259$ queries on average over uniformly random keys.  The two-stage
search is optimal: the components with $e_i > 1$ range over
$B\cdot e_i + 1$ values, namely $141$ and $71$, and
$\lceil\log_2 141\rceil = 8$ and $\lceil\log_2 71\rceil = 7$ match the
counts above, so no query is wasted.  These are information-theoretic
bounds under a perfect oracle; they do not account for hardware fault
rates or repeated-trial overhead.

\paragraph{Cycle dimension.}
The same secret threshold $s_i$ is tested $r = 13$ times per outer
iteration, once for each orientation tuple, and a glitch on any of these
instances yields the same oracle bit. This relaxes the timing precision
the attacker needs, not the per-execution success probability, which the
single-fault model of \Cref{sec:attack:model} fixes at $p$.

\section{Experimental results}
\label{sec:results}
We evaluate the attack
in two stages. The first part is a \emph{software-level fault simulation}, in
which we inject precise faults through flags that select the location of
the fault, and evaluate the resulting outputs. The second stage is a
more \emph{realistic setup} in which we run a hardware experiment on a
ChipWhisperer-Lite (CWLITE-ARM) board; the board's memory and timing
budget does not accommodate a full group-action computation, so we
isolate the targeted control site and demonstrate that the fault is
\emph{feasible in practice}. Throughout, we work within the attacker
model of \Cref{sec:attack:model}.


\subsection{Fault positions}
\label{sec:results:sites}


Because the implementation of~\cite{C:Houben25} executes the class-group
action in a fixed order, a fault in the main loop of
Algorithm~\ref{alg:dcsidh_marc} hits a predictable prime-degree step,
whereas in non-deterministic variants the same secret component is
reached at different times across runs. Within that loop, the most
sensitive site is the assignment on line~\ref{line:boolean} of
Listing~\ref{lst:isog-step}, which computes the lane-selection mask:
faulting it does not corrupt an auxiliary value but changes which lane
takes the next isogeny step.

\lstset{numbers=left, escapeinside={(*@}{@*)}}
\begin{lstlisting}[style=c-elegant,
  basicstyle=\ttfamily\scriptsize, numberstyle=\tiny, numbersep=4pt,
  aboveskip=4pt, belowskip=0pt,
  caption={Action computation on deterministic CSIDH.},label={lst:isog-step}]
uint ell_i;
uint_set(&ell_i, primes[i]);
bool c = (e[i] <= k); (*@\codlinelabel{line:boolean}@*)
proj_cswap(&Aright, &Aleft, c);
proj_cswap(&right_kernel, &left_kernel, c);
proj_cswap(&P, &Q, c);
xMUL(&kernel_point, &Aright, &right_kernel, &cofactor);      /* order primes[i]              */
xISOG(&Aright, &Q, &right_kernel, &kernel_point, primes[i]); /* push Q, right_kernel through */
xMUL(&left_kernel, &Aleft, &left_kernel, &ell_i);            /* both kernels: order cofactor */
proj_cswap(&Aright, &Aleft, c);
proj_cswap(&right_kernel, &left_kernel, c);
proj_cswap(&P, &Q, c);
\end{lstlisting}

\paragraph{OSIDH-LD fault target.}
The same idea applies directly to OSIDH-LD.
The core loop of \texttt{eval\_action} in the reference implementation
mirrors Algorithm~1 of \cite{Houben25_asiacrypt} exactly: for each prime
index~$i$ and inner
step~$t$, the implementation computes a secret-dependent bitmask and
uses it to select the active lane before evaluating the isogeny.
More precisely, the sensitive assignment is given in Listing~\ref{lst:osidh-fault-site}.

\lstset{numbers=left, escapeinside={(*@}{@*)}}
\begin{lstlisting}[style=c-elegant,
  basicstyle=\ttfamily\scriptsize, numberstyle=\tiny, numbersep=4pt,
  aboveskip=4pt, belowskip=0pt,
  caption={Fault target in \texttt{eval\_action} (\texttt{protocols.c}).},
  label={lst:osidh-fault-site}]
digit_t b = compare_gt(t, s[i]); (*@\codlinelabel{line:osidh-b}@*)
swap_vertices(&(R[0]), &(R[1]), b);
/* ... scalar multiplication + isogeny evaluation on R[0] ... */
swap_vertices(&(R[0]), &(R[1]), b);
\end{lstlisting}
Here \texttt{compare\_gt} returns the all-ones word when
$t > s_{\mathit{eff}}$ and zero otherwise, and \texttt{swap\_vertices}
swaps all field elements without
branching. This is the mask of~\Cref{eq:mask}: the first
$s_{\mathit{eff}}$ of the $e_i$ inner steps run on \texttt{R[0]} and the
rest on the companion lane. Forcing \texttt{b} to zero at a chosen
$(u, j, i, t)$ keeps the step on \texttt{R[0]}, misrouting one isogeny
step exactly as in the deterministic CSIDH case.

\subsection{Software-fault analysis}
\label{sec:results:osidh}

We implemented a software fault model on the released OSIDH-LD
reference code~\cite{OSIDH26}.
The implementation provides a drop-in companion function
\texttt{eval\_action\_faulty}, which is identical to
\texttt{eval\_action} except that it intercepts the
\texttt{compare\_gt} call at line~\ref{line:osidh-b} of
Listing~\ref{lst:osidh-fault-site} for a chosen tuple
$(u,\, j,\, i,\, t)$ and forces the result to
zero, leaving the rest of the computation unchanged.
The corresponding faulty public-key function
\texttt{osidhld\_PublicKey\_faulty} wraps this modified action,
tracks which outer iteration is currently executing via a global
counter, and fires the fault exactly once at the specified
coordinates.
This experiment requires no change to the isogeny formulas and
assumes no hardware fault model beyond the ability to corrupt one
secret-dependent mask value.

When compiled without \texttt{-DNDEBUG}, the reference implementation
checks after each outer iteration that the two lanes still encode the
same curve:
\begin{lstlisting}[style=c-elegant,
  basicstyle=\ttfamily\scriptsize, numberstyle=\tiny, numbersep=4pt,
  aboveskip=4pt, belowskip=0pt,
  caption={Lane-consistency assertion in \texttt{eval\_action}.},
  label={lst:osidh-assert}]
#ifndef NDEBUG
    ec_j_inv(&j1, &E1);
    ec_j_inv(&j2, &E2);
    assert(fp2_is_equal(&j1, &j2));
#endif
\end{lstlisting}
It is natural to expect a misrouted step to trip this assertion. Within
the faulted outer iteration it does not. We ran the faulted action with
assertions live, instrumenting the per-$j$ lane comparison: at the
faulted cycle position the two lanes agree, both on $j$-invariants and
on the stronger $A_{24}$ coefficients. The same holds when a single
\texttt{swap\_vertices} is skipped instead of the comparison.

The reason is the meet-in-the-middle structure of the inner loop. Within
cycle position $j$, lane~$0$ starts at $E_j$ and lane~$1$ at $E_{j+1}$,
traversed in the opposite direction, and each $\ell_i$-isogeny is routed
to one lane or the other. Writing $\mathfrak{a}_0,\mathfrak{a}_1$ for
the sub-products sent to the two lanes, the lanes meet precisely when
$\mathfrak{a}_0\mathfrak{a}_1 = \mathfrak{c}$, the fixed class linking
adjacent cycle vertices. A misrouted step at index $i^{*}$ moves that
one factor from the companion lane to lane~$0$:
$\mathfrak{a}_0 \mapsto \mathfrak{a}_0\mathfrak{l}_{i^{*}}$ and
$\mathfrak{a}_1 \mapsto \mathfrak{a}_1\mathfrak{l}_{i^{*}}^{-1}$. The product
is unchanged, so the meeting condition still holds and \emph{both} lanes
arrive at the same curve---namely
$\mathfrak{l}_{i^{*}} * (\text{honest curve})$, the curve the honest
computation would reach with $s_{i^{*}}$ replaced by $s_{i^{*}}+1$. The
invariant the assertion tests is preserved by exactly the perturbation
the attack induces.

This invariance is local to the faulted iteration. Afterwards, position
$j$ holds $\mathfrak{l}_{i^{*}} * E_j$ while its neighbours are honest,
so adjacent vertices are no longer linked by $\mathfrak{c}$. If the
fault occurs before the last outer iteration ($u < B-1$), the next
iteration starts its lanes from these vertices, the lanes at positions
$j-1$ and $j$ end $\mathfrak{l}_{i^{*}}$ apart, and a build with
assertions enabled aborts there. Only a fault in the last outer
iteration ($u = B-1$) passes every check.

Two consequences. First, the implementation's own consistency check is
not a countermeasure against this fault: it is compiled out of release
builds (\texttt{-DNDEBUG}), and even with assertions enabled it misses
every fault in the last outer iteration. Second, every effective fault
is visible in the output: a debug build aborts, and otherwise the
returned orientation cycle differs from the honest one. Comparing all
$r = \mathtt{CYCLE\_LENGTH} = 13$ $j$-invariants of the faulted
orientation cycle against those of a reference run is precisely
the oracle used for recovery; an abort reads as a mismatch.

\paragraph{Full key recovery in software.}
The two-stage binary search of \Cref{sec:attack:osidh} is implemented
on top of this fault oracle.  \texttt{recover\_one\_component} runs the
block search over $u$ at $t = 1$ and dispatches to the inner search over
$t$ only for the two primes with $e_i > 1$; \texttt{recover\_full\_key}
iterates it over all $49$ prime indices and verifies the reconstructed
key against the victim's public key at the end.  Under a perfect output-equality oracle, the run recovers the full key
within the $297$-query worst case of \Cref{sec:attack:osidh}; hardware
fault rates and the repeated-trial overhead are treated below.

\subsection{ChipWhisperer-Lite analysis}
\label{sec:results:dcsidh-chip}
To show that the control bit identified in
\Cref{sec:results:sites} can be faulty on real hardware, we ran
clock-glitching experiments on a \href{https://chipwhisperer.readthedocs.io/en/sca101/Starter\%20Kits/ChipWhisperer-Lite.html}{ChipWhisperer-Lite} platform.
The target is
an STM32F303RCT6 microcontroller: a 32-bit Arm Cortex-M4 running at up to
\SI{72}{\mega\hertz}, with \SI{256}{\kilo\byte} of flash and
\SI{48}{\kilo\byte} of RAM (\SI{40}{\kilo\byte} SRAM plus
\SI{8}{\kilo\byte} CCM). As a compiler we use a 
\texttt{arm-none-eabi-gcc} version $16.2.0$. 

A full class-group action for the larger \csidh parameter sets exceeds the
board's memory and timing budget, so our aim is not to benchmark the complete
protocol but to \emph{reproduce the fault at the sensitive control site and measure
how often it succeeds.} Following~\cite{legrow_24}, we therefore port only the
fault-sensitive fragment to the Cortex-M4. The reliability of that single flip
is what the oracle of our recovery procedure requires, together with the
zero-mask behaviour discussed below. 

More precisely, we target the mask assignment at line~\ref{line:boolean} of
Listing~\ref{lst:isog-step}, aiming for the case where it evaluates to
\texttt{True} in a fault-free run---by \Cref{eq:mask}, a step that the key
routes to the companion lane---but is read as \texttt{False} after a glitch.
On the board, the mask assignment and the first swap are bracketed by
\texttt{trigger\_high()} and \texttt{trigger\_low()}.

Sweeping the glitch parameters, the most productive region used an offset in
$[-21,13]$ and a pulse width of $12$--$20$; within it we observed $3$
successful instruction skips out of $1281$ executions, i.e., an empirical
per-attempt probability of $\approx 2.34\times10^{-3}$, with a $95\%$
Clopper--Pearson interval of roughly $[5\times10^{-4},\, 7\times10^{-3}]$.
The rate is low, but it shows that the control bit can be flipped on a
commodity evaluation setup. The search was not tuned for maximum fault rate,
and more precise equipment---laser injection, or finer clock and voltage
control---would improve it.


\paragraph{Analysis of OSIDH-LD.} We now show, on the same platform, that
the \texttt{compare\_gt} mask can be faulted in practice on the OSIDH-LD
reference implementation~\cite{OSIDH26}.

\paragraph{Target and method.}
As for deterministic CSIDH, the full \SI{256}{\bit} $\mathbb{F}_{p^2}$
group action does not fit the board, so we isolate the
\texttt{compare\_gt} comparison and constant-time swap of
Listing~\ref{lst:osidh-fault-site} in a minimal harness
(Listing~\ref{lst:osidh-trigger}). Between \texttt{trigger\_high()} and
\texttt{trigger\_low()}, it applies them to a two-word proxy state
initialised to distinct sentinels, then reports over SimpleSerial which
sentinel \texttt{R0} holds. We choose $t > s$, so the honest mask is
all-ones (\Cref{eq:mask}) and \texttt{R0} ends with
\texttt{SENTINEL\_R1}, reported as \texttt{0x01}; a successful
$1\!\to\!0$ fault leaves \texttt{SENTINEL\_R0}, reported as
\texttt{0x00}, and any other value is a corrupted response. Since
\texttt{R1} is not read back, \texttt{0x00} shows only that \texttt{R0}
was not overwritten.

\begin{lstlisting}[style=c-elegant,
  basicstyle=\ttfamily\scriptsize, numberstyle=\tiny, numbersep=4pt,
  aboveskip=4pt, belowskip=0pt,
  caption={On-device fault site (\texttt{target\_osidh.c}): the
  \texttt{compare\_gt}\,+\,\texttt{swap} of Listing~\ref{lst:osidh-fault-site},
  isolated on a two-word proxy state and bracketed by the trigger. The sentinel
  check turns the outcome into the one-bit oracle read over SimpleSerial.},
  label={lst:osidh-trigger}]
volatile uint32_t R0 = SENTINEL_R0;   /* 0xAAAAAAAA */
volatile uint32_t R1 = SENTINEL_R1;   /* 0x55555555 */

trigger_high();
digit_t b = compare_gt(t, s);         /* 0xFF..FF iff t > s */
swap_words(&R0, &R1, b);              /* branchless conditional swap */
trigger_low();
/* R0 == SENTINEL_R1 : swap happened  -> 0x01 (correct)        */
/* R0 == SENTINEL_R0 : swap skipped   -> 0x00 (fault injected) */
\end{lstlisting}

We clock-glitch at the nominal \SI{7.37}{\mega\hertz} target clock, inserting a
single glitched cycle (\texttt{scope.glitch.repeat}$=1$) so that at most one
instruction is skipped, and sweep three parameters: the external offset
\texttt{ext\_offset} (which clock cycle after the trigger) and the intra-cycle
\texttt{offset} and \texttt{width} of the pulse.

\begin{figure}[htp]
  \centering
  \begin{subfigure}[b]{0.6\linewidth}
    \centering
    \includegraphics[width=\linewidth]{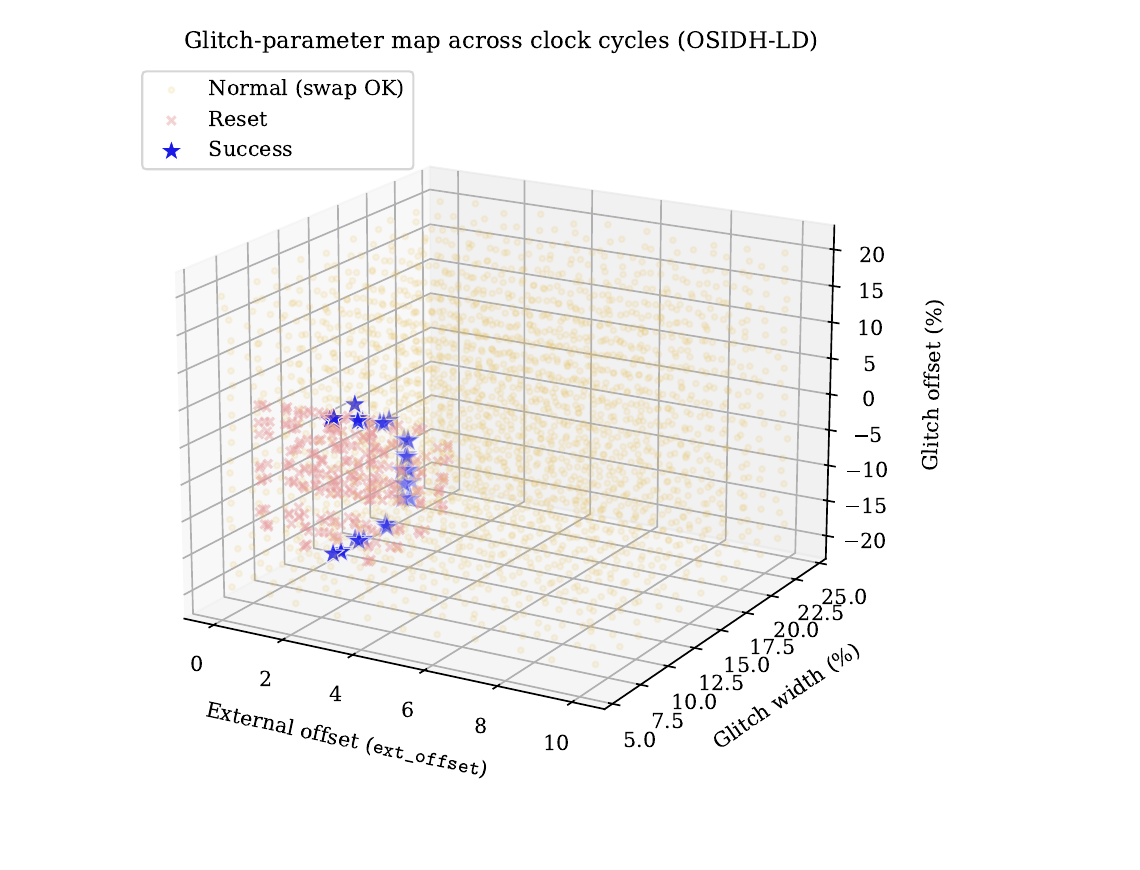}
    \caption{Outcome map over $\mathtt{ext\_offset}\in[0,10]$.}
    \label{fig:osidh-3d}
  \end{subfigure}\hfill
  \begin{subfigure}[b]{0.5\linewidth}
    \centering
    \includegraphics[width=\linewidth]{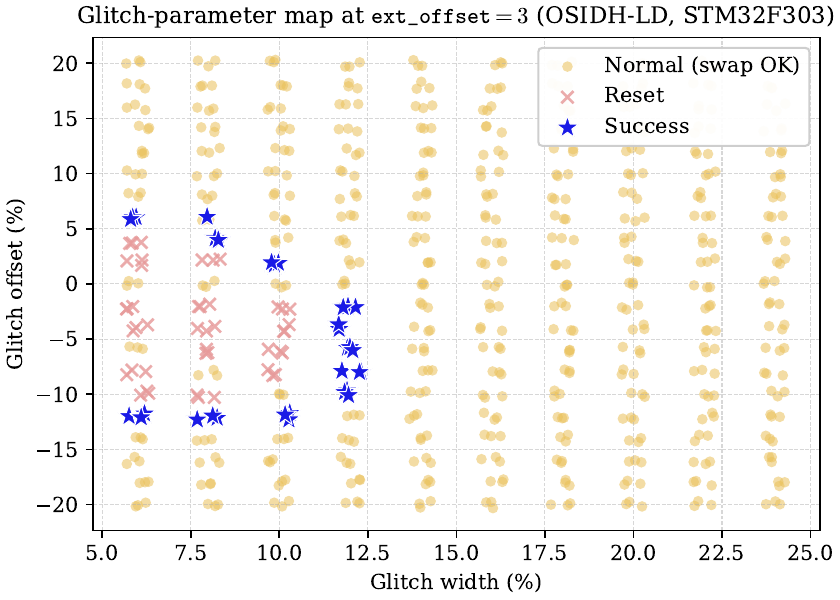}
    \caption{The $\mathtt{ext\_offset}=3$ slab in the
    $(\mathtt{width},\mathtt{offset})$ plane.}
    \label{fig:osidh-coarse}
  \end{subfigure}
  \caption{Coarse clock-glitch sweep on the STM32F303 (single-skip mode). Each
  attempt is a success, with \texttt{R0} left at its own sentinel
  (\emph{Success}), a device reset (\emph{Reset}), or a correct survival
  (\emph{Normal}). (a)~Every success concentrates in a single cycle,
  $\mathtt{ext\_offset}=3$, confirming that the comparison and swap execute
  there. (b)~Eleven points of that slab reach $3/3$ with no resets; we operate
  at $(\mathtt{width},\mathtt{offset})=(12,-6)$.}
  \label{fig:osidh-sweep}
\end{figure}

\paragraph{Fault-parameter search.}
A coarse sweep over $\mathtt{ext\_offset}\in[0,10]$,
$\mathtt{offset}\in[-20,20]$ and $\mathtt{width}\in[6,24]$ (steps $1$,
$2$, $2$; $2310$ points, $3$ attempts each) confines every effective
fault to $\mathtt{ext\_offset}=3$, the cycle in which the comparison and
swap execute (\Cref{fig:osidh-3d}; \Cref{fig:osidh-coarse} shows this
slab in the $(\mathtt{width},\mathtt{offset})$ plane). Eleven points, all
within $\mathtt{width}\in[6,12]$ and $\mathtt{offset}\in[-12,6]$, reach
$3/3$ with no resets or corrupted responses, and only one edge point
drops to $1/3$. The fault is thus clean and tolerant of timing jitter; we
take $(\mathtt{ext\_offset},\mathtt{offset},\mathtt{width})=(3,-6,12)$
as the operating point.

\paragraph{Fault reliability.}
At the operating point, all $50$ consecutive attempts returned
\texttt{0x00} with no resets, giving $p \geq 0.94$ (one-sided $95\%$
Clopper--Pearson bound) at the isolated site. The deterministic-CSIDH
rate of $3/1281 \approx 2.34\times10^{-3}$ is not directly comparable,
since that sweep was not first localised to a single cycle; what this
figure shows is that, once the glitch is aligned to the comparison, the
skip is essentially deterministic.

\begin{figure}[htp]
  \centering
  \includegraphics[width=0.8\linewidth]{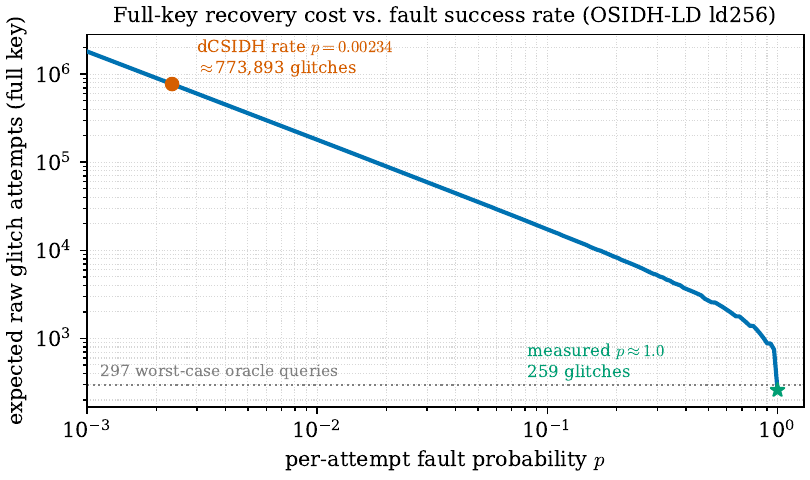}
  \caption{Expected raw glitches for full-key recovery on OSIDH-LD
  (\texttt{ld256}) as a function of the per-attempt fault probability $p$, under
  the repeated-trial model (per-threshold miss probability $\le 10^{-6}$). The
  dotted line is the $297$-query worst case; the average over all components is
  $259$. The markers show the dCSIDH rate and the measured OSIDH-LD operating
  point. The model assumes the glitch has no effect on steps whose honest mask
  is zero.}
  \label{fig:osidh-budget}
\end{figure}

\paragraph{From fault rate to recovery cost.}
Recovery compares the faulted public key against the honest one over all
$r=13$ $j$-invariants of the orientation cycle and binary-searches each of the
$n=49$ secret components, an information-theoretic worst case of $297$ oracle
queries (\Cref{sec:attack:osidh}). Turning queries into physical attempts
depends on $p$: an ``effective'' outcome confirms $s_i \ge \tau$ on the first
observed skip, whereas a ``no-op'' outcome must be confirmed by enough repeats
that a genuine skip would almost surely have appeared---the repeated-trial
overhead noted by~\cite{legrow_24}. \Cref{fig:osidh-budget} plots the resulting
expected number of raw glitches for a full key against $p$. At the measured
$p \approx 1$ the no-op confirmation collapses to one attempt per threshold,
so recovery costs about $259$ glitches on average ($297$ worst case)---one per
query; at the lower confidence bound $p = 0.94$ it costs about $753$. At the
dCSIDH rate $p = 3/1281$ the same model predicts about $7.7\times10^{5}$
attempts, and between $2.6\times10^{5}$ and $3.8\times10^{6}$ across its
Clopper--Pearson interval. These are estimates under the fault model of
\Cref{sec:attack:model}, not measured attack costs: they assume that the
glitch leaves the output unchanged on steps whose honest mask is zero (see
below).

\noindent
\textbf{Scope and limitations.}
Both hardware campaigns fault only steps whose honest mask is all-ones
($t > s$). Recovery also reads queries on zero-mask steps as no-ops,
relying on the model of \Cref{sec:attack:model} that a $1 \to 0$ fault
on a zero bit has no effect. 

As for deterministic CSIDH, we fault an isolated control site behind a
dedicated trigger, not a full on-device group action. A complete
OSIDH-LD execution evaluates the comparison
$B \cdot r \cdot \sum_i e_i = 35 \cdot 13 \cdot 53 = 24\,115$ times, in
groups of $r = 13$ equivalent instances per target $(u,i,t)$. What
remains is \emph{trigger isolation}, firing the already reliable skip on
the chosen $(u,i,t)$, which the deterministic, public schedule should in
principle allow from a single power trace.

\section{A fault-resistant variant}

With the attack presented in this paper as evidence, we argue that a fault-protected implementation of \csidh must not be deterministic. Indeed, a deterministic implementation
lets the attacker choose which isogeny computation to fault. This is
consistent with~\cite{EC:BKLMPR23}, where the disorientation attack
required fewer samples against CTIDH than against the original \csidh.
We therefore propose countermeasures that break determinism while
keeping the schemes constant-time and dummy-free. By introducing
randomness, we aim to prevent the output of a faulted execution from
immediately revealing information about a specific secret component
$s_i$. We analyse two countermeasures: shuffling the order of the primes,
and randomizing the order of the outer loop. Shuffling
suffices when a fault causes an execution error, as in OSIDH-LD,
provided one extra iteration with the all-zero exponent vector is
appended so that no fault lands in the unchecked last iteration. In
deterministic \csidh, a fault yields a valid but wrong curve that cannot
be detected at little extra cost, so the second countermeasure is needed
there.

\subsection{Randomizing the order of the primes}
In Algorithms~\ref{alg:dcsidh_marc} and~\ref{alg:osidh_ld}, the prime-degree isogenies are computed in a fixed and publicly known order. In the OSIDH-LD case, for each $i$, the algorithm computes $e_i$ consecutive isogenies of degree $\ell_i$, processing the primes in increasing index order. Therefore, a fault targeting the secret-dependent control bit at loop indices $u$, $i$, $j$ and $t$ can be directly associated with the prime $\ell_i$, and hence with the secret threshold $s_i$. As shown in Section~\ref{sec:attack}, the attacker can recover the targeted control bit and hence the lane on which the $\ell_i$-isogeny is computed: the default lane if the bit is $0$, and the companion lane otherwise.

The order in which the prime-degree isogenies are computed does not affect the final output. This follows from the commutativity of the underlying class-group action. Therefore, instead of processing the primes in the fixed order $1,2,\ldots,n$, we propose sampling a random permutation of the prime indices and evaluating the corresponding prime-degree isogenies according to this randomized order. A fresh permutation is sampled for every new execution of the algorithm. As a result, the attacker no longer knows in advance which prime-degree isogeny is processed at a given physical loop position and therefore loses the ability to deliberately target a chosen $s_i$. Nevertheless, forcing the secret-dependent control bit from $1$ to $0$ still causes the corresponding prime-degree isogeny to be evaluated on the wrong lane.

In what follows, we analyse the effectiveness of this countermeasure in both the deterministic \csidh case and the OSIDH-LD case. \emph{We consider a countermeasure effective not only when it prevents full key recovery, but also when it significantly increases the number of successful faults required by the attack.}

\paragraph{Protecting \csidh.}

For the proposed non-deterministic implementation of \csidh given in Algorithm~\ref{alg:ndcsidh_marc}, a precisely timed fault forcing the secret-dependent control bit from $1$ to $0$ causes the current prime-degree isogeny to be computed on the wrong lane. For a fault targeting the isogeny computation corresponding to outer-loop counter $j$ and inner-loop counter $k$, one $\ell_i$-isogeny is computed on the default lane instead of the companion lane, where $i := \tau(k)$. At a high level, this amounts to forcing the algorithm to apply the action of the ideal class corresponding to the exponent vector
\[
(s_1,\ldots,s_i+1,\ldots,s_n).
\]
Since this still corresponds to a valid computation, but with an incorrect key, the algorithm returns a valid curve. In particular, the faulty output $E_f$ and the correct output $E_c$ satisfy
\[
E_c = [\mathfrak{l}_i]^{-2} \ast E_f.
\]
This corresponds to a disorientation fault and, even though the degree $\ell_i$ of the corrupted isogeny is not known, the attacker can determine it by brute-forcing all $n$ possibilities.

However, if the secret-dependent control bit is already $0$, the fault does not alter the normal execution, and the attacker observes the correct output. Therefore, to recover the control bit for a given prime $\ell_i$, the attacker must repeatedly inject faults in a fixed loop position until either a corrupted output differing from the correct one by an $\ell_i^2$-isogeny is observed, or until the probability is high enough that $\tau(k)=i$ has occurred at least once. In the first case, the attacker concludes that the corresponding control bit is $1$, otherwise, the attacker concludes with high probability that it is $0$.

Using this observation, one can design a simple attack against Algorithm~\ref{alg:ndcsidh_marc}. Namely, for each $j=1,\ldots,B$, the attacker repeatedly injects faults at the fixed loop position $(j,k)$, until $\tau(k)$ has taken every value in $\{1,\ldots,n\}$ at least once with high enough probability. From the outputs that differ from the correct one, the attacker learns the primes $\ell_i$ for which the corresponding control bit was $1$, and therefore $s_i<j$. For the remaining primes, the attacker can assume with high probability that $s_i\geq j$. The number of faults required for each fixed fault position $(j,k)$ corresponds to the well-known coupon collector problem~\cite{AnceaumeBSS16,coupon1}. Since $\tau$ is sampled uniformly at random for every execution, $\tau(k)$ is uniformly distributed over $\{1,\ldots,n\}$. Therefore, the expected number of successful faults required is
\[
nH_n
=
n\sum_{k=1}^{n}\frac{1}{k},
\]
where $H_n$ is the $n$-th harmonic number. Repeating this for all $B$ values of $j$ gives that the expected total number of required successful faults to recover all $s_i$ is $B \cdot nH_n$.

As in Section~\ref{sec:attack:recovery}, binary search reduces the
number of successful faults, but in this case, it yields a more elaborate analysis. For each prime $\ell_i$,
Algorithm~\ref{alg:ndcsidh_marc} computes the $s_i$ isogenies on the default lane
before the $B-s_i$ on the companion lane, so the control bits of each prime form
a non-decreasing sequence and $s_i$ is the position of the rightmost $0$
($s_i=0$ if all bits are $1$). We run the binary searches for all primes
in parallel, grouping the primes by their current search interval,
initially $[1,B]$. In the first step, every prime is queried at
$M=\lfloor(1+B)/2\rfloor$, leaving the interval $[1,M-1]$ or $[M+1,B]$;
each later step queries every group at the midpoint of its interval,
doubling the number of groups, and the search ends after at most
$\lceil\log_2(B+1)\rceil$ steps.

Parallelism pays off because all primes in a group share the same
midpoint. As discussed earlier, $nH_n$ successful faults are expected to
determine the control bit at $j=\lfloor(1+B)/2\rfloor$ for all primes. If this splits the
primes into groups of sizes $m$ and $n-m$, the next two queries need
$nH_m$ and $nH_{n-m}$ faults in expectation, by the variant of the
coupon collector problem in which each coupon in a fixed set of $m$ out of $n$
coupons must be drawn at least once.

To get a feeling for how this parallel binary search attack performs compared to the one discussed in Section~\ref{sec:attack:recovery}, we first consider a scenario in which the required number of successful faults is expected to get close to the theoretical \emph{upper} bound. Namely, at each step of the binary search with search interval $[L,R]$, the corresponding group of primes splits almost perfectly into two equal subgroups associated with the subintervals $[L,M-1]$ and $[M+1,R]$. This means that the parallel binary search visits all leaves of the tree representing the binary search, and the number of primes associated with each leaf is as equal as possible. This achieves the theoretical upper bound assuming that $B=2^x-1$ for some $x>0$, i.e., the tree is a perfect binary tree, and $n$ is divisible by $(B+1)/2$. Then, under this assumption, the expected total number of faults $T$ required to recover all secret components $s_i$ is given by the simplified formula.

\[
T =nH_n
 + 2nH_{\frac{n}{2}}
 + 4nH_{\frac{n}{4}}
 + 8nH_{\frac{n}{8}}
 + \cdots
 + \tfrac{(B+1)}{2} n H_{\frac{2n}{B+1}} 
 = n\sum_{k=0}^{x-1} 2^k H_{\frac{n}{2^k}}.
\]

Although no closed-form expression is available in the general case, this special case gives a useful benchmark. The attack from Section~\ref{sec:attack:recovery} requires at most $n\left\lceil\log_2(B+1)\right\rceil$ faults. We then observe that
\[
n\sum_{k=0}^{x-1}2^kH_{\frac{n}{2^k}}
\geq
n\sum_{k=0}^{x-1}2^k
=
n(2^x-1)
=
nB
>
n\left\lceil\log_2(B+1)\right\rceil.
\]
Thus, under the assumptions stated above, the countermeasure increases the number of faults required for full key recovery.

When the assumptions of $B$ being an almost-power-of-two and $(B+1)/2$ dividing $n$ are not met, this formula overestimates the worst case. For instance, for the parameters of deterministic CSIDH-512 ($n=74$ and $B=5$), the formula (taking $x=\left\lceil \log_2 B \right\rceil = 3$) gives us an approximation of 2025 successful faults, while the correct count is 1605. We obtained the latter by finding the worst-case key using dynamic programming techniques, since we can not derive a closed formula.

On the other extreme, we consider a scenario in which the required number of successful faults is close to the theoretical \emph{lower} bound. First, we observe that
\[
H_m+H_{n-m}>H_n
\]
for all $1\leq m<n$. Hence, splitting a group of primes into two non-empty subgroups increases the combined expected cost of the next binary search step. Using this observation, one can show that the lower bound is achieved when all primes follow the same sequence of steps in the binary search, which corresponds to $s_1=\cdots=s_n$. In this case, the expected number of successful faults required per step is $nH_n$. Thus, for fixed parameters $B$ and $n$, the expected total number of faults required for full key recovery is approximately

\[
nH_n\log_2(B+1).
\]
In fact, under the prescribed binary search, the minimum possible number of steps is $\left\lfloor\log_2(B+1)\right\rfloor$, while the maximum is $\left\lceil\log_2(B+1)\right\rceil$, depending on the value that all $s_i$ take. For the CSIDH-512 parameter set, the minimum possible number of steps performed by the binary search is
\[
\left\lfloor \log_2(B+1) \right\rfloor
=
\left\lfloor \log_2 6 \right\rfloor
=
2.
\]
Therefore, the expected number of successful faults required by the parallel binary search is approximately bounded below by
\[
74 \cdot H_{74} \cdot \left\lfloor \log_2 6 \right\rfloor
\approx
74 \cdot 4.888 \cdot 2
\approx 723.
\]

The countermeasure significantly increases the fault complexity of full key recovery. More importantly, it renders the attack probabilistic, i.e., the exponent threshold is lower-bounded only with some probability even under the assumption of perfect fault injection success rate. The candidate solution can ultimately be corrected with a meet-in-the-middle search to the non-faulty curve. However, each mistake induces an exponential growth in the complexity of this final phase of the key recovery. There is also a natural trade-off between the number of faults used for the attack (going beyond the coupon collector problem threshold) and the correctness offset of the obtained key candidate.

\begin{algorithm}[htp]
\caption{Non-deterministic CSIDH. Highlighted lines randomize the order of the primes.}
\label{alg:ndcsidh_marc}
\KwIn{Same as Algorithm~\ref{alg:dcsidh_marc}.}
\KwOut{Same as Algorithm~\ref{alg:dcsidh_marc}.}

$(P', Q') \gets (P, Q)$\;
\cmprime{$\tau \gets \text{a random permutation from } S_n$}\;
\For{$j \gets 1$ \KwTo $B$}{
    $(P^+, P^-) \gets (P', Q')$\;
    $m \gets \prod_i \ell_i$\;

    \For{$k \gets 1$ \KwTo $n$}{
        \cmprime{$i \gets \tau(k)$}\;
        \eIf{$j \leq s_i$}{
            $m \gets m/\ell_i$\;
            $K \gets [m]P^+$\;
            $(\Ep, P^+, Q') \gets \evaliso(\Ep, K, P^+, Q')$\;
            $P^- \gets [\ell_i]P^-$\;
        }{
            $m \gets m/\ell_i$\;
            $K \gets [m]P^-$\;
            $(\Em, P^-, P') \gets \evaliso(\Em, K, P^-, P')$\;
            $P^+ \gets [\ell_i]P^+$\;
        }
    }

    $(\Em, P') \gets \evalfiso(\Em, P')$\;
    \textbf{assert} $\Ep = \Em$\;
}

\Return{$(E^+, P', Q')$}\;
\end{algorithm}

\paragraph{Protecting OSIDH-LD.}
In Algorithm~\ref{alg:osidh_ld_random}, the $j$-th output tuple is computed using the $j$-th and $(j+1)$-th tuples from the input full kernel representation, independently of the other tuples in the output full kernel representation. Therefore, a fault affecting one outer-loop iteration corrupts only the corresponding output tuple.

The effect of the fault is similar to the deterministic \csidh case. For a fixed outer-loop counter $j$, forcing the secret-dependent control bit from $1$ to $0$ shifts the $j$-th output curve by one $\ell_i$-isogeny in the positive direction:
\[
E_j^{\mathrm{correct}}
=
[\mathfrak{l}_i]^{-1} * E_j^{\mathrm{faulted}},
\]
where $i := \tau(k)$ denotes the prime index selected by the randomized ordering. More precisely, the fault causes the meeting point of the two lanes to shift. Since the two lanes still reach a common codomain, the lane-consistency checks during the same iteration of the group action do not detect the fault. The result is therefore a full kernel representation in which one tuple is internally consistent,  but the required relation between this tuple and its neighboring tuples is no longer satisfied:
\[
\prod_{i=1}^{n} [\mathfrak{l}_i]^{e_i} * E_{j-1}
=
E_j
=
\prod_{i=1}^{n} [\mathfrak{l}_i]^{-e_i} * E_{j+1}.
\]
This becomes relevant when the resulting representation is fed as input to the next group-action iteration. During the next application of Algorithm~\ref{alg:osidh_ld_random}, when processing the same outer-loop counter $j$, the two lanes fail to meet, and the lane-consistency checks detect this unexpected behavior of the algorithm.

For the fault-resistance analysis, we assume that the algorithm aborts whenever a lane-consistency check fails. Consider a fault targeting outer iteration $u<B-1$ of the group action, counting from $0$ as in Section~\ref{sec:attack:osidh}. If the targeted control bit is $0$, the attacker obtains the fault-free result and cannot recover the degree $\ell_i$ of the corresponding isogeny computation. If the control bit is $1$, the algorithm instead aborts with an error, and the attacker still cannot determine the degree of the affected isogeny.

The caveat is that, for faults targeting the last iteration ($u=B-1$), the lane-consistency checks will not cause the algorithm to abort. Therefore, in the case where the control bit is $1$, the attacker observes the corrupted output and can infer the prime degree of the targeted isogeny. More precisely, if $s_i > (B-1)e_i$,
then the attacker can recover $s_i$. Ignoring for simplicity the two primes that have $e_i > 1$, the required number of successful faults is approximately $nH_n = 49 H_{49} \approx 219$. To mitigate this issue and make the class-group action resistant to the proposed attack, we suggest adding one additional group-action iteration at the end that applies the action of the ideal class corresponding to the all-zero exponent vector. This additional step does not change the final output, but protects the secret components $s_i$. For the \texttt{ld256} parameter set, this corresponds to using $B = \mathtt{ITERATIONS} = 36$ applications of Algorithm~\ref{alg:osidh_ld_random}. We note that the trade-off between time complexity and security is justifiable in this case, since the final variant of the algorithm fully protects the secret key against the proposed attack.

OSIDH-LD comes with several algorithmic optimizations compared to the original variant in~\cite{Houben25_asiacrypt}. The analysis of the impact of our randomization techniques on these optimizations, as well as their eventual reconstruction, remains open.

\begin{algorithm}[htp]
\caption{OSIDH-LD group action with randomized prime ordering (one iteration). Highlighted lines randomize the order of the primes.}
\label{alg:osidh_ld_random}
\KwIn{Same as Algorithm~\ref{alg:osidh_ld}.}
\KwOut{Same as Algorithm~\ref{alg:osidh_ld}.}

\cmprime{$\tau \gets \text{a random permutation from } S_n$}\;

\For{$j \gets 1$ \KwTo $r$}{
    $R_0 \gets (E_j,\; P_j,\; Q_j)$\;
    $R_1 \gets (E_{j+1},\; Q_{j+1},\; P_{j+1})$\tcp*{indices mod $r$}
    $m \gets \prod_{i=1}^{n} \ell_i^{\,e_i}$\;

    \For{$k \gets 1$ \KwTo $n$}{
        \cmprime{$i \gets \tau(k)$}\;
        \For{$t \gets 1$ \KwTo $e_i$}{
            \tcp{Perform the corresponding $\ell_i$-isogeny step as in Algorithm~\ref{alg:osidh_ld}}
        }
    }

    $R_0[1] \gets R_1[2]$\;
    $E'_j \gets \IsoFix(R_0,\; R_1)$\tcp*{isomorphism identification for $E'_j$}
}
\Return{$E'_1, \ldots, E'_r$}\;
\end{algorithm}

\subsection{Randomizing the order of the outer loop}
To further strengthen the countermeasure in the CSIDH case, we propose an additional modification to Algorithm~\ref{alg:ndcsidh_marc}, namely to randomize the order of execution of the outer loop. In Algorithm~\ref{alg:ndcsidh_marc}, for each prime we first compute the $s_i$ isogenies on the default lane, and then we compute the $B-s_i$ isogenies on the companion lane. This is a crucial characteristic that lets us upper/lower bound $s_i$ with one query and allows for a binary search. With this countermeasure, we still compute $s_i$ isogenies on the default lane and $B-s_i$ on the companion one, but in a randomized order. In this case, when a disorientation fault shows that a given $\ell_i$ was computed on the wrong lane, the only information we can infer is that $s_i<B$.

This countermeasure is not leakage-free, because an attacker can still try to perform a statistical analysis and potentially reduce the search space for a meet-in-the-middle-style attack. We briefly summarize the ideas, without giving a concrete estimation on the number of required successful faults, mainly because it is hard to say how close a key candidate must be for the final search to be practical.

One option is for the attacker to fault only a single physical loop position $(h,k)$ for fixed $h$ and $k$. Then, the pair $(j,i)$, where $j:=\rho(h)\in\{1,\ldots,B\}$ and $i:=\tau(k)\in\{1,\ldots,n\}$, can take $B\cdot n$ possible values. We use the observation that if one knows the proportion
\[
R_i := \frac{B-s_i}{B}
\]
of isogenies on the companion lane among the total number of isogenies, then the secret value $s_i$ can be recovered as
$
s_i=(1-R_i)\cdot B.
$
We therefore approximate $R_i$ from the observations obtained after injecting sufficiently many faults at the fixed loop position $(h,k)$. Assume that a total of $T$ faults is injected, and let $T_i$ denote the number of times a faulty output corresponding to the prime $\ell_i$ is observed. Since such an output implies that the targeted control bit was $1$, and the degree of the corresponding isogeny can be recovered by brute force, $T_i$ can be determined from the faulty outputs. We can then use the following estimator for $R_i$:
\[
R_i' := \frac{T_i}{T/n},
\]
where $T/n$ represents the average number of faults associated with each prime, since the primes are sampled uniformly at random. Note that this estimate is imprecise because the collected data only tells us how many times the prime $\ell_i$ appeared on the companion lane, but not how many times it appeared in total.
Nevertheless, a candidate for the secret key is then the exponent vector $(s_1',\ldots,s_n')$, where
$
s_i' := (1-R_i')\cdot B.
$

\begin{algorithm}[htp]
\caption{Even more non-deterministic CSIDH. Lines highlighted in blue randomize the order of the primes, as in Algorithm~\ref{alg:ndcsidh_marc}; lines highlighted in orange randomize the order of the outer loop.}
\label{alg:ndcsidh_2}
\KwIn{Same as Algorithm~\ref{alg:dcsidh_marc}.}
\KwOut{Same as Algorithm~\ref{alg:dcsidh_marc}.}

\cmprime{$\tau \gets \text{a random permutation from } S_n$}\;
\cmloop{$\rho \gets \text{a random permutation from } S_B$}\;

\For{$h \gets 1$ \KwTo $B$}{
    \cmloop{$j \gets \rho(h)$}\;
    $(P^+,P^-) \gets (P',Q')$\;
    $m \gets \prod_{i=1}^{n}\ell_i$\;

    \For{$k \gets 1$ \KwTo $n$}{
        \cmprime{$i \gets \tau(k)$}\;

        \eIf{$j \leq s_i$}{
            $m \gets m/\ell_i$\;
            $K \gets [m]P^+$\;
            $(E^+,P^+,Q') \gets
            \evaliso(E^+,K,P^+,Q')$\;
            $P^- \gets [\ell_i]P^-$\;
        }{
            $m \gets m/\ell_i$\;
            $K \gets [m]P^-$\;
            $(E^-,P^-,P') \gets
            \evaliso(E^-,K,P^-,P')$\;
            $P^+ \gets [\ell_i]P^+$\;
        }
    }

    $(\Em, P') \gets \evalfiso(\Em, P')$\;
    \textbf{assert} $\Ep = \Em$\;
}

\Return{$(E^+, P', Q')$}\;
\end{algorithm}


\subsection*{Acknowledgements}
We thank Daniel Koynov for developing a script for the upper-bound estimate for the attack on CSIDH with prime-order randomization.
\subsection*{AI statement}
We used an AI assistant to check grammar and correct typos in some sections. 
The Python script was written with the help of
an AI assistant; the authors reviewed the code and verified its output.

\bibliographystyle{plain}
\bibliography{bibliography}

\begin{thebibliography}{10}

\bibitem{AdjCR20}
Gora Adj, Jes{\'{u}}s{-}Javier Chi{-}Dom{\'{\i}}nguez, and Francisco
  Rodr{\'{\i}}guez{-}Henr{\'{\i}}quez.
\newblock On new v{\'{e}}lu's formulae and their applications to {CSIDH} and
  {B-SIDH} constant-time implementations.
\newblock {\em {IACR} Cryptol. ePrint Arch.}, page 1109, 2020.

\bibitem{AnceaumeBSS16}
Emmanuelle Anceaume, Yann Busnel, Ernst Schulte{-}Geers, and Bruno Sericola.
\newblock Optimization results for a generalized coupon collector problem.
\newblock {\em J. Appl. Probab.}, 53(2):622--629, 2016.

\bibitem{BanegasBCCLMSS21}
Gustavo Banegas, Daniel~J. Bernstein, Fabio Campos, Tung Chou, Tanja Lange,
  Michael Meyer, Benjamin Smith, and Jana Sot{\'{a}}kov{\'{a}}.
\newblock {CTIDH:} faster constant-time {CSIDH}.
\newblock {\em {IACR} Trans. Cryptogr. Hardw. Embed. Syst.}, 2021(4):351--387,
  2021.

\bibitem{BanegasHS25}
Gustavo Banegas, Andreas Hellenbrand, and Matheus Saldanha.
\newblock Hardened {CTIDH:} dummy-free and deterministic {CTIDH}.
\newblock In Ratna Dutta, Luca~De Feo, and Sugata Gangopadhyay, editors, {\em
  Progress in Cryptology - {INDOCRYPT} 2025 - 26th International Conference on
  Cryptology in India, Bhubaneshwar, India, December 14-17, 2025, Proceedings},
  volume 16372 of {\em Lecture Notes in Computer Science}, pages 194--215.
  Springer, 2025.

\bibitem{EC:BKLMPR23}
Gustavo Banegas, Juliane Kr{\"{a}}mer, Tanja Lange, Michael Meyer, Lorenz
  Panny, Krijn Reijnders, Jana Sot{\'{a}}kov{\'{a}}, and Monika Trimoska.
\newblock Disorientation faults in {CSIDH}.
\newblock In Carmit Hazay and Martijn Stam, editors, {\em Advances in
  Cryptology - {EUROCRYPT} 2023 - 42nd Annual International Conference on the
  Theory and Applications of Cryptographic Techniques, Lyon, France, April
  23-27, 2023, Proceedings, Part {V}}, volume 14008 of {\em Lecture Notes in
  Computer Science}, pages 310--342. Springer, 2023.

\bibitem{secsidh}
Fabio Campos, Jorge Ch{\'{a}}vez{-}Saab, Jes{\'{u}}s{-}Javier
  Chi{-}Dom{\'{\i}}nguez, Michael Meyer, Krijn Reijnders, Francisco
  Rodr{\'{\i}}guez{-}Henr{\'{\i}}quez, Peter Schwabe, and Thom Wiggers.
\newblock Optimizations and practicality of high-security {CSIDH}.
\newblock {\em {IACR} Commun. Cryptol.}, 1(1):5, 2024.

\bibitem{CamposHMR25}
Fabio Campos, Andreas Hellenbrand, Michael Meyer, and Krijn Reijnders.
\newblock {dCTIDH:} fast {\&} deterministic {CTIDH}.
\newblock {\em {IACR} Trans. Cryptogr. Hardw. Embed. Syst.}, 2025(3):516--541,
  2025.

\bibitem{CamposKMOS20}
Fabio Campos, Matthias~J. Kannwischer, Michael Meyer, Hiroshi Onuki, and Marc
  St{\"{o}}ttinger.
\newblock Trouble at the {CSIDH:} protecting {CSIDH} with dummy-operations
  against fault injection attacks.
\newblock In {\em 17th Workshop on Fault Detection and Tolerance in
  Cryptography, {FDTC} 2020, Milan, Italy, September 13, 2020}, pages 57--65.
  {IEEE}, 2020.

\bibitem{CamposKM21}
Fabio Campos, Juliane Kr{\"{a}}mer, and Marcel M{\"{u}}ller.
\newblock Safe-error attacks on {SIKE} and {CSIDH}.
\newblock In Lejla Batina, Stjepan Picek, and Mainack Mondal, editors, {\em
  Security, Privacy, and Applied Cryptography Engineering - 11th International
  Conference, {SPACE} 2021, Kolkata, India, December 10-13, 2021, Proceedings},
  volume 13162 of {\em Lecture Notes in Computer Science}, pages 104--125.
  Springer, 2021.

\bibitem{Castryck0MPR18}
Wouter Castryck, Tanja Lange, Chloe Martindale, Lorenz Panny, and Joost Renes.
\newblock {CSIDH:} an efficient post-quantum commutative group action.
\newblock In Thomas Peyrin and Steven~D. Galbraith, editors, {\em Advances in
  Cryptology - {ASIACRYPT} 2018 - 24th International Conference on the Theory
  and Application of Cryptology and Information Security, Brisbane, QLD,
  Australia, December 2-6, 2018, Proceedings, Part {III}}, volume 11274 of {\em
  Lecture Notes in Computer Science}, pages 395--427. Springer, 2018.

\bibitem{EC:CasPanVer20}
Wouter Castryck, Lorenz Panny, and Frederik Vercauteren.
\newblock Rational isogenies from irrational endomorphisms.
\newblock In Anne Canteaut and Yuval Ishai, editors, {\em Advances in
  Cryptology - {EUROCRYPT} 2020 - 39th Annual International Conference on the
  Theory and Applications of Cryptographic Techniques, Zagreb, Croatia, May
  10-14, 2020, Proceedings, Part {II}}, volume 12106 of {\em Lecture Notes in
  Computer Science}, pages 523--548. Springer, 2020.

\bibitem{Cervantes-Vazquez19}
Daniel Cervantes{-}V{\'{a}}zquez, Mathilde Chenu, Jes{\'{u}}s{-}Javier
  Chi{-}Dom{\'{\i}}nguez, Luca~De Feo, Francisco
  Rodr{\'{\i}}guez{-}Henr{\'{\i}}quez, and Benjamin Smith.
\newblock Stronger and faster side-channel protections for {CSIDH}.
\newblock In Peter Schwabe and Nicolas Th{\'{e}}riault, editors, {\em Progress
  in Cryptology - {LATINCRYPT} 2019 - 6th International Conference on
  Cryptology and Information Security in Latin America, Santiago de Chile,
  Chile, October 2-4, 2019, Proceedings}, volume 11774 of {\em Lecture Notes in
  Computer Science}, pages 173--193. Springer, 2019.

\bibitem{Chavez-SaabCJR20}
Jorge Ch{\'{a}}vez{-}Saab, Jes{\'{u}}s{-}Javier Chi{-}Dom{\'{\i}}nguez, Samuel
  Jaques, and Francisco Rodr{\'{\i}}guez{-}Henr{\'{\i}}quez.
\newblock The {SQALE} of {CSIDH:} square-root v{\'{e}}lu quantum-resistant
  isogeny action with low exponents.
\newblock {\em {IACR} Cryptol. ePrint Arch.}, page 1520, 2020.

\bibitem{Chi-DominguezR22}
Jes{\'{u}}s{-}Javier Chi{-}Dom{\'{\i}}nguez and Francisco
  Rodr{\'{\i}}guez{-}Henr{\'{\i}}quez.
\newblock Optimal strategies for {CSIDH}.
\newblock {\em Adv. Math. Commun.}, 16(2):383--411, 2022.

\bibitem{legrow_24}
Tinghung Chiu, Jason LeGrow, and Wenjie Xiong.
\newblock Practical fault injection attacks on constant time csidh and
  mitigation techniques.
\newblock In {\em Proceedings of the 2024 Workshop on Attacks and Solutions in
  Hardware Security}, ASHES '24, page 11–22, New York, NY, USA, 2024.
  Association for Computing Machinery.

\bibitem{C:Houben25}
Marc Houben.
\newblock Deterministic algorithms for class group actions.
\newblock In Yael~Tauman Kalai and Seny~F. Kamara, editors, {\em Advances in
  Cryptology - {CRYPTO} 2025 - 45th Annual International Cryptology Conference,
  Santa Barbara, CA, USA, August 17-21, 2025, Proceedings, Part {I}}, volume
  16000 of {\em Lecture Notes in Computer Science}, pages 100--130. Springer,
  2025.

\bibitem{Houben25_asiacrypt}
Marc Houben.
\newblock Efficient post-quantum commutative group actions from orientations of
  large discriminant.
\newblock In Goichiro Hanaoka and Bo{-}Yin Yang, editors, {\em Advances in
  Cryptology - {ASIACRYPT} 2025 - 31st International Conference on the Theory
  and Application of Cryptology and Information Security, Melbourne, VIC,
  Australia, December 8-12, 2025, Proceedings, Part {IV}}, volume 16248 of {\em
  Lecture Notes in Computer Science}, pages 141--173. Springer, 2025.

\bibitem{legrow_20}
Jason LeGrow and Aaron Hutchinson.
\newblock An analysis of fault attacks on {CSIDH}.
\newblock Cryptology {ePrint} Archive, Paper 2020/1006, 2020.

\bibitem{MeyerCR19}
Michael Meyer, Fabio Campos, and Steffen Reith.
\newblock On lions and elligators: An efficient constant-time implementation of
  {CSIDH}.
\newblock In Jintai Ding and Rainer Steinwandt, editors, {\em Post-Quantum
  Cryptography - 10th International Conference, PQCrypto 2019, Chongqing,
  China, May 8-10, 2019 Revised Selected Papers}, volume 11505 of {\em Lecture
  Notes in Computer Science}, pages 307--325. Springer, 2019.

\bibitem{OnukiAYT20}
Hiroshi Onuki, Yusuke Aikawa, Tsutomu Yamazaki, and Tsuyoshi Takagi.
\newblock A constant-time algorithm of {CSIDH} keeping two points.
\newblock {\em {IEICE} Trans. Fundam. Electron. Commun. Comput. Sci.},
  103-A(10):1174--1182, 2020.

\bibitem{IWSEC:OnuTak20}
Hiroshi Onuki and Tsuyoshi Takagi.
\newblock On collisions related to an ideal class of order 3 in {CSIDH}.
\newblock In Kazumaro Aoki and Akira Kanaoka, editors, {\em Advances in
  Information and Computer Security - 15th International Workshop on Security,
  {IWSEC} 2020, Fukui, Japan, September 2-4, 2020, Proceedings}, volume 12231
  of {\em Lecture Notes in Computer Science}, pages 131--148. Springer, 2020.

\bibitem{OSIDH26}
Weize Wang, Yi-Fu Lai, Kaizhan Lin, and Yunlei Zhao.
\newblock Efficient and parallel implementation of isogeny-based deterministic
  group actions.
\newblock {\em IACR Transactions on Cryptographic Hardware and Embedded
  Systems}, 2026(3):541–565, Jul. 2026.

\bibitem{coupon1}
Weiyu Xu and Ao~Kevin Tang.
\newblock A generalized coupon collector problem.
\newblock {\em CoRR}, abs/1010.5608, 2010.

\end{thebibliography}



\end{document}